\documentclass[aps,twocolumn,groupedaddress,showpacs]{revtex4}
\usepackage{epsf,latexsym}
\usepackage{times}

\newcommand{\bra}[1]{\langle #1 |}
\newcommand{\ket}[1]{| #1 \rangle}

\newcommand\R{{\mathrm {I\!R}}}

\newcommand\h{{\cal H}}

\newcommand{\ra}{{\rightarrow}}
\newcommand{\be}{\begin{equation}}
\newcommand{\ee}{\end{equation}}

\def\CC{{\rm\kern.24em \vrule width.04em height1.46ex depth-.07ex
    \kern-.30em C}}
\def\P{{\rm I\kern-.25em P}}

\def\bbbone{{\mathchoice {\rm 1\mskip-4mu l} {\rm 1\mskip-4mu l}
{\rm 1\mskip-4.5mu l} {\rm 1\mskip-5mu l}}} 

\def\bbbc{{\mathchoice {\setbox0=\hbox{$\displaystyle\rm C$}\hbox{\hbox
to0pt{\kern0.4\wd0\vrule height0.9\ht0\hss}\box0}}
{\setbox0=\hbox{$\textstyle\rm C$}\hbox{\hbox
to0pt{\kern0.4\wd0\vrule height0.9\ht0\hss}\box0}}
{\setbox0=\hbox{$\scriptstyle\rm C$}\hbox{\hbox
to0pt{\kern0.4\wd0\vrule height0.9\ht0\hss}\box0}}
{\setbox0=\hbox{$\scriptscriptstyle\rm C$}\hbox{\hbox
to0pt{\kern0.4\wd0\vrule height0.9\ht0\hss}\box0}}}}

\def\bbbz{{\mathchoice {\hbox{$\sf\textstyle Z\kern-0.4em Z$}}
{\hbox{$\sf\textstyle Z\kern-0.4em Z$}}
{\hbox{$\sf\scriptstyle Z\kern-0.3em Z$}}
{\hbox{$\sf\scriptscriptstyle Z\kern-0.2em Z$}}}}

\newcommand{\putfig}[2]{$$\leavevmode\hbox{\epsfxsize=#2 cm
   \epsffile{#1.eps}}$$}

\begin{document}
\title{Quantum information processing in bosonic lattices}
\author{Radu Ionicioiu}
\author{Paolo Zanardi}
\affiliation{Istituto Nazionale per la Fisica della Materia (INFM), UdR Torino-Politecnico, 10129 Torino, Italy}
\affiliation{Institute for Scientific Interchange (ISI), Villa Gualino, Viale Settimio Severo 65, I-10133 Torino, Italy}

\begin{abstract}
We consider a class of models of self-interacting bosons hopping on a lattice. We show that properly tailored
space-temporal coherent control of the single-body coupling parameters allows for universal quantum 
computation in a given sector of the global Fock space. This general strategy for encoded universality 
in bosonic systems has in principle several candidates for physical implementation.
\end{abstract}

\maketitle

The central problem in quantum information processing (QIP) \cite{qip} is the ability to control a quantum 
system in order to achieve some predefined purpose like quantum computation (QC). In general, a quantum information 
processor is realized by assembling a large number of copies of a given quantum 
system e.g., a qubit, and by making these copies interact in a controlled coherent fashion.
A crucial issue of any proposal for a QIP implementation is its scalability i.e., the 
realizability, at least in principle, of the above structure for an arbitrary large size. An appropriate 
architecture to achieve this last goal is naturally provided by a lattice with sites hosting the processing quantum systems. 

In this paper we shall describe a general scheme for performing quantum computation with interacting 
bosonic particles in a lattice. The system under study is very general 
and several existing QC proposals could fit into this framework. These include optical 
qubits \cite{milburn, simple_qc, cerf}, Josephson junction qubits \cite{jj} and optical lattice loaded with 
ultracold bosonic atoms e.g., Bose-Einstein condensates (BECs) \cite{bec_lattice, vp, bec_qubit}. Without 
focusing on any of these particular implementations, we will develop a general framework for encoding qubits  
and for performing universal quantum gates on such encoded qubits. Despite their highly practical relevance, we 
will not discuss decoherence issues since they are strongly dependent on the specific physical implementation.

Let us start by casting the problem we are going to address in a more precise control-theoretic fashion.
The single mode Fock space will be denoted by $h:=\mbox{span}\{|n\rangle\}_{n=0}^\infty$. The Hamiltonian 
acting on ${\cal H}_{\Lambda}:= h^{\otimes\,L}$ that we would like to analyze is given by:
\be
H({\cal V})= \sum_{i,j\in\Lambda} (V^{(2)}_{ij} n_i\,n_j + V^{(1)}_{ij} c_i^\dagger c_j+ \mbox{H.c.}) 
\label{Ham}
\ee 
where:
(i) $\Lambda$ is an index set (the lattice vertices) with $L$ elements;
(ii) $c_j,\,c_j^\dagger (j\in\Lambda)$ are bosonic creation and annihilation operators and $n_i:=c_i^\dagger c_i$ 
the corresponding occupation numbers;
(iii) ${\cal V} :=\{V^{(2)}_{ij}\}\times \{V^{(1)}_{ij}\}\subset \R \times \CC,$
is the set of quasi-classical 'control' parameters.
The Hamiltonian (\ref{Ham}) represents a generalized Bose-Hubbard model \cite{BH}, the terms weighted by 
the $V^{(2)}_{ij}$'s account for the non-linear two-body interactions whereas the $V^{(1)}_{ij}$'s are one-body 
terms describing the hopping of the bosonic particles among the lattice sites. The interplay of these two terms is 
known to give rise to a rich quantum phase-diagram with insulating and superfluid regions \cite{bec_lattice,bec_mott}.

The ultimate goal is to find an $M$-qubit encoding $e: (\CC^2)^{\otimes\,M} \mapsto {\cal H}_\Lambda$ such that 
control on the parameters $V_{ij}(t)$ in (\ref{Ham}) would enact universal computational capabilities on the code.
In this paper we will propose such an encoding which will enable us to perform universal quantum computation on a suitable 
sector of the Fock space associated with the bosonic lattice \cite{Fock}. 
It is worthwhile to keep in mind that even though we will consider only the "spatial" interpretation of the single-particle 
modes i.e., spatially localized wave-functions, they could even be momentum modes or modes associated to any other 
single-particle wavefunction.

\begin{figure}
\putfig{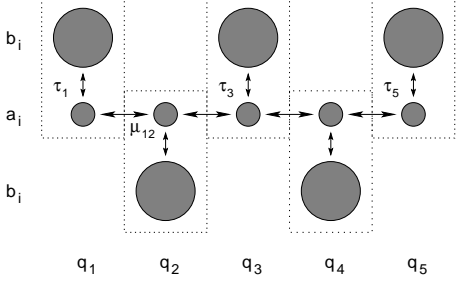}{6}
\caption{An example of a qubit array. The $a$ ($b$) modes correspond to the small (big) disks; the $b$ modes are situated in 
a {\em zigzag} geometry in order to minimize their interaction. Inter-qubit coupling is given only via the $a$ modes.}
\label{array}
\end{figure}

\noindent {\it The qubit.} We define the qubit using {\em two} lattice sites (dual rail encoding). We denote 
by $a_i (a_i^\dagger)$ and $b_i (b_i^\dagger)$ the corresponding annihilation (creation) operators for the two 
bosonic modes ($i$ is the qubit index). The Hamiltonian of the system has two terms: $H= H_0 + H_{int}$. The first 
term is the sum of all single-qubit Hamiltonians:
\begin{eqnarray}
\nonumber
H_0= \sum_i H_i =\sum_i \varepsilon_{1,i}\, n_{a,i}^2+ \varepsilon_{2,i}\, n_{b,i}^2+ \\ 
\, +\gamma_{1,i}\, n_{a,i}+ \gamma_{2,i}\, n_{b,i} + \tau_i (a_i^\dagger\, b_i + a_i\, b_i^\dagger)
\label{h0}
\end{eqnarray}
with $n_{a,i}= a_i^\dagger\, a_i$, $n_{b,i}= b_i^\dagger\, b_i$; $\tau_i$ is the tunneling rate between the $a$ and $b$ modes 
of the {\em same} qubit ({\em intra}-qubit tunneling rate). The second term represents the interaction between different qubits:
\be
H_{int}= \sum_{i\neq j} \mu_{ij} (a_i^\dagger\, a_j + a_i\, a_j^\dagger)+ \sum_{i\neq j} \chi_{ij}\, n_{a,i}\, n_{a,j}
\label{hint}
\ee
We assume that qubits interact only via the $a$ modes (with $\mu_{ij}$ the {\em inter}-qubit tunneling rate; $\chi_{ij}$ is the 
Kerr coupling). An intuitive picture is given in Fig.~\ref{array}. Another possible geometry, for example, is to have a common 
bus to which only the $a$-mode of each qubit is coupled \cite{bur}.

We define the logical state of the qubit by the number of particles in the $a$ mode. Thus, $n_a= 0$ defines the logical 
$\ket{0}_{\sf L}$ state and $n_a= 1$ the logical $\ket{1}_{\sf L}$ state. The computational space is therefore restricted 
to $n_a \in \{ 0, 1 \}$ and we will show how to enforce this condition after each gate operation. It is important to stress that the 
subsystems which support our qubits are (finite-dimensional subspaces) of bosonic modes rather than particles. Therefore, the paradigm 
we adopt here concerning quantum entanglement for systems of indistinguishable particles is the one advocated in ref.~\cite{fermi}. We describe the main steps of any quantum computation scheme: universal set of gates, state preparation and measurement.

The following set of gates is universal for quantum computation \cite{gates}: $\{ \sf H, P_\varphi, C_\pi \}$, where 
${\sf H}= \frac{1}{\sqrt{2}}\pmatrix{1&1 \cr 1&-1}$ is a Hadamard gate, ${\sf P}_\varphi = \mbox{diag}\,(1,\, e^{i\varphi})$ 
is a single-qubit phase shift, and $\sf C_\pi$ is a controlled sign flip. We use the more general controlled phase gate 
${\sf C_\varphi} = \mbox{diag}\, (1,\,1,\,1,\,e^{i\varphi})$. We enact these gates by controlling the time dependence of the parameters characterizing the system. For the single-qubit gates we set all $\mu_{ij}= \chi_{ij}= 0$ and we vary only the single-qubit parameters $\varepsilon_{\alpha,i}(t)$, $\gamma_{\alpha,i}(t)$ and $\tau_i(t)$ ($\alpha = 1, 2$). On the contrary, for the two-qubit gate, we keep constant the single-qubit parameters and we 
control only the inter-qubit tunneling rate $\mu_{ij}(t)$ or the Kerr coupling $\chi_{ij}(t)$.\\
{\em Single-qubit gates.} In the absence of any external coupling ($\mu_{ij}= \chi_{ij}= 0$), the single-qubit Hamiltonian 
is (for simplicity we omit the qubit index $i$):
\be
H_1= \varepsilon_1\, n_a^2+ \varepsilon_2\, n_b^2+ \gamma_1\, n_a+ \gamma_2\, n_b + \tau (a^\dagger b + a b^\dagger)
\label{h1}
\ee
The total particle number $n= n_a+ n_b$ is conserved, since $[H, n]=0$. Thus, the one-qubit Hilbert space splits into a direct 
sum $\h = \bigoplus_{n=0}^\infty \h_n$, as the Hamiltonian (\ref{h1}) leaves invariant the subspaces $\h_n$ with total particle 
number $n= \rm const$ and ${\rm dim}\, \h_n= n+1$. In view of this decomposition of the Hilbert space, we can relabel the Fock 
states as $\ket{n_a\, n_b}\equiv \ket{n;\, n_a}$. Thus, any vector can be written as $\ket{\psi} = \sum_n\sum_{i=0}^n c_{n,i}\, 
\ket{n;\, i}$. Since our initial state will be a Fock state and since the Hamiltonian conserves the total particle number $n$, 
the single-qubit wave-function at any time will always remain in the subspace $\h_n$, $\ket{\psi(t)} = \sum_{i=0}^n c_{n,i}(t) 
\ket{n;\, i} \in \h_n.$ Basically, the only degree of freedom left is the number of particles 
$i\, (= n_a)$ in the $a$-mode.

\begin{figure}
\putfig{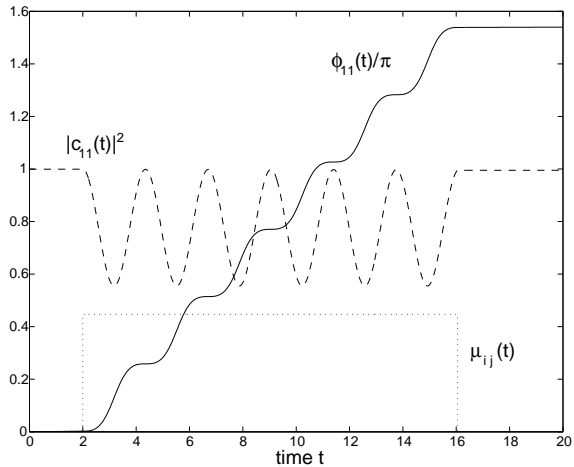}{7.5}
\caption{Amplitude and phase (in $\pi$ units) of $\ket{11}$ state during the $\sf C_\varphi$ gate operation. The dynamical phase 
$e^{-2i\varepsilon t}$ has been subtracted.}
\label{c2}
\end{figure}

For $\tau=0$, the Fock states $\ket{n; i}$ are are also energy eigenvectors, with eigenvalues given by (we omit the label $n$ since the total number of particle is conserved):
\be
E_i \equiv \bra{n; i}\, H\, \ket{n; i}= \varepsilon_1 i^2 + \varepsilon_2 (n-i)^2+ \gamma_1 i+ \gamma_2 (n-i)
\ee
$i= 0 \ldots n.$ We also want our qubit states (defined as before 
$\ket{0}_{\sf L}\equiv \ket{n; 0}$, $\ket{1}_{\sf L}\equiv \ket{n; 1}$) to correspond to the degenerate ground state 
$E_0= E_1$. This implies the following energy degeneracy condition:
\be
\varepsilon_1- (2n-1)\varepsilon_2 + \gamma_1- \gamma_2=0
\label{e0e1}
\ee
For the Hadamard gate, we keep $\varepsilon_k(t)$, $\gamma_k(t)$ constant (and satisfying the degeneracy 
condition (\ref{e0e1})) and we allow only a time-dependent tunneling rate $\tau(t)$. Since the energy gap between the ground and 
the first excited state is $\Delta E\equiv E_2- E_1= 4n \varepsilon_2+ 2(\gamma_2- \gamma_1)= 2(\varepsilon_1+ \varepsilon_2)$, we can treat the system in a first 
approximation as a degenerate two-level system, ignoring higher level transitions. The time evolution (up to a phase) is given by 
the operator $U(t)= e^{-i\sigma_x \int_0^t \tau(t')dt'}$, equivalent to a rotation around the $x$-axis 
${\sf R_x}(\theta)\equiv  e^{-i\theta\sigma_x}$. Then we can obtain the Hadamard gate as $\sf H= P_{\pi/2}\, R_x(\pi/4)\, P_{\pi/2}$. 
Similarly, we have ${\sf NOT}= i\,{\sf R_x(\pi/2)}$. In order to confirm this simple analysis, we have performed a full time-dependent 
simulation in the whole Hilbert space. We numerically integrate the Hamiltonian (\ref{h1}) for $n=30$. If we adiabatically switch 
the tunneling rate $\tau(t)$, we can control the population of higher levels (and hence the leakage form the computational space) to 
be negligible (in our simulation, this is less than $10^{-3}$ for a Gaussian pulse shape $\tau(t)$). The time scale required for performing 
single qubit gates is about one order of magnitude smaller than the one necessary for the two-qubit gate (see Fig.~\ref{c2}).

To enact the phase shift gate $\sf P_\varphi$ we keep the (intra-qubit) tunneling zero (therefore we always stay in the 
computational space) and we allow only a time dependence for (some of) the other parameters $\varepsilon_k(t)$, $\gamma_k(t)$, 
$k=1,2$. Since $\tau= 0$, the time-dependent Hamiltonian is diagonal and we can solve the model analytically. Let $[0,\, T]$ be the 
time interval during which the gate acts. The time evolution of a Fock state is ($\hbar=1$): $\ket{n;\, i} \rightarrow 
e^{-i T \overline E_i}\, \ket{n;\, i}$, with $\overline E_i\equiv \frac{1}{T}\int_0^T E_i(t) dt$ the average value of $E_i(t)$ 
during the gate operation. Then the gate action on the basis states is:
\be
U_\varphi = e^{-i T \overline E_0}\ \mbox{diag}\, (1, e^{-i\varphi})
\ee
with $\varphi= T(\overline E_1- \overline E_0)$. In order to have $\varphi \ne 0$ we need to violate the energy degeneracy 
condition (\ref{e0e1}) by varying any of the four parameters $\varepsilon_k(t)$, $\gamma_k(t)$. From an experimental point of 
view, the self-interactions $\varepsilon_k$ might be harder to control, since they are related to the collision rates (in a BEC, 
for example). On the other hand, $\gamma_k$ are related to the energy offset of the trapping potential and are conceivably  easier 
to control. Thus, we can keep constant any three of these parameters and control only the time variation of the remaining one (say 
$\gamma_1$). This method gives us considerable freedom in choosing the shape and duration of the pulses $\gamma_1(t)$ (the function 
$\gamma_1(t)$ is not even necessary to be continuous, it should be only integrable). Basically, the only condition is 
$\gamma_1(0)= \gamma_1(T)= \gamma_2- \varepsilon_1+ (2n-1)\varepsilon_2$, such that the two-qubit states are again degenerate after the 
gate; this ensures that the phase difference between $\ket{0}_{\sf L}$ and $\ket{1}_{\sf L}$ is 'frozen'.

It is important to note that both rotation angles $\theta$ and $\varphi$ characterizing the single qubit gates depend only on the 
average values of $\tau(t)$ and $\gamma(t)$, respectively, and therefore they are relatively robust under small fluctuations of the 
control parameters (but they vary linearly with the gate time).

\noindent {\em Two-qubit gate.} An important question is: What type of interactions, together with the one-qubit gates discussed 
previously, are universal? We will discuss two kind of couplings, both nonlinear, which achieve this.\\
{\bf (i)} $H_{ij}^K = \chi_{ij} n_{a,i}\, n_{a,j}$. This is the well-known Kerr Hamiltonian and is used for optical qubits to enact 
${\sf C}_\varphi$ \cite{milburn}. However, in usual materials the nonlinearity (the so-called $\chi^{(3)}$) is a few orders of 
magnitudes smaller than what is needed, and hence this scheme for producing the two-qubit gate is impractical. The Hamiltonian can 
be easily integrated and the gate action on a two-qubit state is simply given by $U= \mbox{diag}\, (1, 1, 1, e^{-iT\chi_{ij} })$, 
since in our dual rail encoding we always have $n_{a,i}, n_{a,j} \in \{0,1\}$. We note that, by considering excitons in semiconductor quantum dots as bosons \cite{hardcore}, this nonlinearity is the one used to enact the two-qubit gate in the QIP proposal of Biolatti {\it et al.} \cite{Bio}.

In the following we analyze in more detail a second universal (along with the one-qubit gates) Hamiltonian.\\
{\bf (ii)} $H_{ij} = \varepsilon (n_{a,i}^2+ n_{a,j}^2)+ \gamma (n_{a,i}+ n_{a,j})+ \mu_{ij} (a_i^\dagger\, a_j + a_i\, a_j^\dagger)$. 
This is the Hamiltonian of two-qubits $i$ and $j$ interacting via the $a$-modes. It is identical to the one-qubit Hamiltonian 
(\ref{h1}), with $\varepsilon_{1,i}= \varepsilon_{1,j}= \varepsilon$ and $\gamma_{1,i}= \gamma_{1,j}= \gamma$, but now we also have 
$n_{a,i}, n_{a,j} \in \{0,1\}$. Since the total number of particles is conserved $[H_{ij}, n_{a,i}+ n_{a,j}]= 0$, we can neglect the 
constant term proportional to $\gamma$ and rewrite the Hamiltonian as (with the obvious notation 
$n_i\equiv n_{a,i}$, $n_j\equiv n_{a,j}$)
\be
H_{ij}(t) = \varepsilon (n_i^2+ n_j^2)+ \mu_{ij}(t) (a_i^\dagger\, a_j + a_i\, a_j^\dagger)
\label{hij}
\ee
Given the Hamiltonian (\ref{hij}), we want to find the control parameter $\mu_{ij}(t)$ such that the action of the gate is:
\be
\ket{n_i\, n_j}\ \ra\ e^{i\varphi_{n_i n_j}} \ket{n_i\, n_j}
\label{ninj}
\ee
This ensures that the total particle number for each qubit is conserved {\em after} the gate operation, i.e., there is no leakage 
from the computational space. Of course, {\em during} the gate operation this is not true, since intermediate states like $\ket{02}$ 
do not correspond to any logical state, but we will cancel these unwanted states dynamically. Again, let $[0,\, T]$ be the time 
interval during which the gate acts. There are three possible cases, depending on the initial state. Since the logical state 
$\ket{n_i\, n_j}_{\sf L}$ is the same as the Fock state $\ket{n_i\, n_j}$, we can omit the subscript ${\sf L}$ (keeping in mind that 
some {\em intermediate} states will not correspond to any logical state). The possible input states belong to different 
representations of $H_{ij}$ with total particle number $n=0,1,2$, respectively ($n\equiv n_i+ n_j$).\\
{\bf (a)} $\ket{00}$. This case is trivial, $\ket{00}\ \ra\ \ket{00}$.\\
\noindent
{\bf (b)} $\ket{01}$ and $\ket{10}$. For $n= 1$, we have $H_{ij}^{(1)}(t)= \varepsilon\bbbone + \mu_{ij}(t)\sigma_x$ (the 
superscript (1) refers to the total particle number $n$). Since $[H_{ij}^{(1)}(t_1), H_{ij}^{(1)}(t_2)]=0$ at all times, we can 
analytically integrate the time evolution to obtain:
\be
U^{(1)}(T)= e^{-i\varepsilon T}(\bbbone \cos\omega_1 T -i \sigma_x \sin\omega_1 T)
\label{10}
\ee
with $\omega_1= \overline \mu_{ij}= \frac{1}{T}\int_0^T \mu_{ij}(t) dt$. Imposing condition (\ref{ninj}), we require 
$\sin(\omega_1 T)= 0$. Therefore $\omega_1 T= m_1 \pi$, $m_1\in \bbbz$. This implies the following transformation for the basis states:
\be
\ket{01}\ \ra\ (-1)^{m_1}e^{-i\varepsilon T} \ket{01}
\ee
and similarly for $\ket{10}$.\\
{\bf (c)} $\ket{11}$. For $n=2$, the Hamiltonian is:
\be
H_{ij}^{(2)}(t)= \pmatrix{4\varepsilon& \sqrt{2}\mu_{ij}(t) & 0 \cr \sqrt{2}\mu_{ij}(t)& 2\varepsilon &\sqrt{2}\mu_{ij}(t)\cr 0& 
\sqrt{2}\mu_{ij}(t) & 4\varepsilon}
\label{hij2}
\ee
In general we cannot integrate this analytically. If $\mu_{ij}(t)= \rm const$, the exact time evolution is
\begin{eqnarray}
\nonumber
\ket{11}\ \ra \ e^{-3i\varepsilon t} \left\{ \ket{11}\left( \cos \omega_2 t + \frac{i\varepsilon}{\omega_2}\sin \omega_2 t \right)- 
\right. \\
\left. -\frac{i\mu_{ij}\sqrt{2}}{\omega_2} \sin\omega_2 t\, ( \ket{02} + \ket{20} ) \right\}
\label{11}
\end{eqnarray}
with $\omega_2= \sqrt{\varepsilon^2+ 4\mu_{ij}^2}$. Again, since we want to recover the $\ket{11}$ state after the gate operation, we 
impose the condition $\sin(\omega_2 T)= 0$, hence $\omega_2 T= m_2 \pi$, $m_2\in \bbbz$. In this case, the evolution of the state 
is $\ket{11}\ \ra\ (-1)^{m_2}e^{-3i\varepsilon T}\ket{11}$. Together with the previous condition (i.e., $\omega_1 T= m_1 \pi$), we obtain
\be
\frac{m_2}{m_1}= \frac{\omega_2}{\omega_1} = \sqrt{\frac{\varepsilon^2}{\mu_{ij}^2} +4}
\ee
Modulo single-qubit phases ${\sf P}_\theta^{\otimes 2}$, $\theta\equiv \pi(m_1- \sqrt{m_2^2-4m_1^2})$, the gate operation on the 
basis states $\ket{n_i\, n_j}$ is equivalent to
\be
U(T)= \mbox{diag}(1, 1, 1, e^{i\phi_{11}})\equiv \sf C_{\phi_{11}}
\ee
with $\phi_{11}\equiv \pi(m_2- \sqrt{m_2^2- 4m_1^2})$. In Fig.~\ref{c2} we present a full time-dependent simulation for the 
evolution of the state $\ket{11}$ with $m_1=2$ and $m_2=6$. We choose a step function for the tunneling rate 
$\mu_{ij}(t)$. The simulation is in good agreement with the exact solution for constant tunneling presented above. After extracting 
the dynamical $e^{-i\varepsilon t}$ phase for {\em each} qubit, the $\ket{11}$ state picks up a phase 
$\phi_{11}/\pi= 6- 2\sqrt{5}\approx 1.53$, whereas the $\ket{01}$ and $\ket{10}$ states remain phaseless.

There is an important point to note here: the nonlinear term $\varepsilon (n_i^2+ n_j^2)$ in (\ref{hij}) is essential for enacting 
the gate. If $\varepsilon= 0$ (or if we replace it with a linear one $\gamma_i\, n_i+ \gamma_j\, n_j$), it can be shown that the gate 
would be equivalent to $\bbbone$. This result is not surprising. The single-body operators entering our basic 
Hamiltonian (\ref{Ham}) i.e., $\{c_i^\dagger c_j\}_{i,j=1}^L$ span, by commutation, a ${\cal A}_L\cong u(L)$ Lie algebra having 
$\hat N:=\sum_{i=1}^L c^\dagger_i c_i$ as central element.
The lattice Fock space splits into $N+L-1 \choose L-1$-dimensional invariant sectors of ${\cal A}_L$
labelled by the eigenvalues $N$ of $\hat N$ i.e., by the total number of bosons.
In order to universally manipulate $M$ encoded qubits by using just the ${\cal A}_L$-elements one has to use 
$L\sim 2^{M}$ lattice sites. On the other hand our encoding scales {\em linearly} with the lattice size i.e., $M=L/2$.
Thus, the nonlinear term in the Hamiltonian (\ref{Ham}) provides an exponential reduction of resources.

\noindent {\it Preparation and measurement.} It is enough to prepare the $\ket{0\, 0 \ldots 0}_{\sf L}$ state of the qubit array. 
We start by preparing two linear optical lattices in which the $b$-mode bosons are held in a zigzag fashion in order to 
minimize their interaction (see Fig.~\ref{array}).
The next step is to create the middle row in Fig.~\ref{array} where the $a$-modes for all qubits will be held. This can be done by
engineering the confining potential in order to create a second minimum for the $a$-modes. At this stage
there is no tunneling between any of these wells, $\tau_i=\mu_{ij}= 0$, and therefore all qubits are in the $\ket{0}_{\sf L}$
state ($n_{a,i}=0$). Another possibility is to start from a Mott insulator phase, in which exact numbers of atoms are localized
at individual lattice sites; this has been recently demonstrated experimentally \cite{bec_mott}.

The measurement technique is conceptually simple -- we have to detect, for each qubit, the presence or the absence of one boson in the $a$-mode. For an optical lattice, this can be done by fluorescence:
an atom present will fluoresce under the right laser illumination. The middle row of the qubit array will be a succession of
dark (bright) spots, i.e., the atom is absent (present) in the $a$-mode, corresponding to qubit in state $\ket{0}_{\sf L}$
($\ket{1}_{\sf L}$).

In conclusion, we have provided a further example of the paradigm of the so-called encoded-universality \cite{eu}.
A limited i.e., non-universal set of controllable interactions can still provide a full computational power in a suitable
encoding subspace. We have presented a general framework for performing encoded universal QC on systems of self-interacting bosonic 
particles hopping on a lattice. Our strategy requires the ability to control in space and time the one-body couplings of the system. A summary of the parameter dependence for the gate operations is shown in Table \ref{tab1} (only one of the two nonlinear interactions $\mu_{ij}$ and $\chi_{ij}$ are sufficient and therefore they can be used alternatively, depending on the system). Possible implementations of this scheme include optical qubits, Josephson junctions and BEC in optical lattices.
\begin{table}
\caption{A minimal example of time dependence for the control parameters. The last line is the degeneracy
condition (\ref{e0e1}).}
\begin{ruledtabular}
\begin{tabular}{l|ccr}
 & ${\sf H}$ & ${\sf P_\varphi}$ & ${\sf C_\varphi}$ \\
\hline
$\varepsilon_{1,i};\, \varepsilon_{2,i}$ & const & const & const, $\ne$0 \\
$\gamma_{1,i}$ & const & $\gamma_{1,i}(t)$ & const \\
$\gamma_{2,i}$ & const & const & const \\
$\tau_i$ & $\tau_i(t)$ & 0 & 0 \\
$\mu_{ij}$ & 0 & 0 & $\mu_{ij}(t)$ \\
$\chi_{ij}$ & 0 & 0 & $\chi_{ij}(t)$ \\
$E_0-E_1$ & 0 & $\ne$0 & 0 \\
\end{tabular}
\end{ruledtabular}
\label{tab1}
\end{table}

\noindent {\bf Acknowledgments.} We are grateful to Mario Rasetti and Paolo Giorda for useful comments.
Special thanks are due to Vittorio Penna for introducing us to the problem of bosonic lattices
and providing a constructive remark for the realization of the one-qubit phase gate.


\begin{thebibliography}{}

\bibitem{qip} D.P.~DiVincenzo and C.~Bennett, Nature {\bf 404}, 247 (2000).

\bibitem{milburn} G.~J.~Milburn, \prl {\bf 62}, 2124 (1989).

\bibitem{simple_qc} I.~L.~Chuang, and Y.~Yamamoto, \pra {\bf 52}, 3489 (1995).

\bibitem{cerf} N.~J.~Cerf, C.~Adami, and P.~G.~Kwiat, \pra {\bf 57}, 1477 (1998); quant-ph/9706022.

\bibitem{jj} Y.~Makhlin, G.~Sch\"on, and A.~Shnirman, \rmp {\bf 73}, 357 (2000).

\bibitem{bec_lattice} D.~Jaksch {\it et al.}, \prl {\bf 81}, 3108 (1998).

\bibitem{vp} R.~Franzosi, V.~Penna, and R.~Zecchina, Int.~J.~Mod.~Phys.~B {\bf 14}, 943 (2000);
R.~Franzosi, and V.~Penna, \pra {\bf 63}, 043609 (2001).

\bibitem{bec_qubit} Z.B.~Chen, and Y.D.~Zhang, \pra {\bf 65}, 022318 (2002).

\bibitem{BH} M.~P.~A.~Fisher {\it et al.}, \prb {\bf 40}, 546 (1989).

\bibitem{bec_mott} M.~Greiner {\it et al.}, Nature {\bf 415}, 39 (2002).

\bibitem{Fock} For massive particles the Fock space itself is not a physical state space (due to mass super-selection rules which forbid superpositions of vectors with different number of particles). In this case for any physical encoding $e( (\CC^2)^{\otimes M})$ must be included in a subspace with fixed particle number.

\bibitem{bur} Bosons on "lattices" with a complex structure have been studied e.g., R.~Burioni {\it et al}, Europhys.~Lett. {\bf 52}, 251 (2000).

\bibitem{fermi} P.~Zanardi, \pra {\bf 65}, 042101 (2002); P.~Zanardi, X-G.~Wang, {\em Fermionic entanglement in itinerant systems}, quant-ph/0201028.

\bibitem{gates} A.~Barenco {\it et al.}, \pra {\bf 52}, 3457 (1995).

\bibitem{hardcore} Excitonic creation operators $X_i$, being fermionic bilinear, satisfy $X_i^2=0$ (hard-core condition). They are called {\em pseudo} bosons.

\bibitem{Bio} E.~Biolatti {\it et al.}, \prl {\bf 85} 5647 (2000); \prb {\bf 65}, 075306 (2002).

\bibitem{eu} J.~Kempe {\it et al.}, \pra {\bf 63}, 042307 (2001); D.~P.~DiVincenzo et al, Nature {\bf 408}, 339 (2000).

\end{thebibliography}
\end{document}